\documentclass[10pt,conference]{IEEEtran}
\usepackage{amsmath,amssymb}
\usepackage{graphicx}
\usepackage{booktabs}

\PassOptionsToPackage{bookmarks=false}{hyperref} % no bookmark pane
\usepackage[hidelinks]{hyperref}

\usepackage[
  top=0.75in,          % was 0.647–0.65; now safely ≥ 0.70
  bottom=1in,
  left=0.68in,
  right=0.68in,
  columnsep=0.22in     % keeps the OK gutter
]{geometry}

\usepackage{cite}
\usepackage{amsfonts}
\usepackage{algorithmic}
\usepackage{textcomp}
\usepackage{enumitem}
\usepackage{tabularx}

\usepackage[table,dvipsnames]{xcolor}
\definecolor{lightgreen}{RGB}{220,255,220}
\definecolor{lightred}{RGB}{255,220,220}
\definecolor{lightgray}{gray}{0.90}

\def\BibTeX{{\rm B\kern-.05em{\sc i\kern-.025em b}\kern-.08em
    T\kern-.1667em\lower.7ex\hbox{E}\kern-.125emX}}
\begin{document}

\title{MX-AI: Agentic Observability and Control Platform for Open and AI-RAN}

\author{\IEEEauthorblockN{Ilias Chatzistefanidis}
\IEEEauthorblockA{\textit{Communications Department} \\
\textit{EURECOM}\\
Sophia Antipolis, France \\
ilias.chatzistefanidis@eurecom.fr}
\and
\IEEEauthorblockN{Andrea Leone}
\IEEEauthorblockA{\textit{AI Department} \\
\textit{BubbleRAN}\\
Sophia Antipolis, France \\
andrea.leone@bubbleran.com}
\and
\IEEEauthorblockN{Ali Yaghoubian}
\IEEEauthorblockA{\textit{Communications Department} \\
\textit{EURECOM}\\
Sophia Antipolis, France \\
ali.yaghoubian@eurecom.fr}
\and
\IEEEauthorblockN{Mikel Irazabal}
\IEEEauthorblockA{\textit{RIC Department} \\
\textit{BubbleRAN}\\
Sophia Antipolis, France \\
mikel.irazabal@bubbleran.com}
\and
\IEEEauthorblockN{Sehad Nassim}
\IEEEauthorblockA{\textit{Info \& Com Engineering} \\
\textit{Aalto University}\\
Espoo, Finland \\
nassim.sehad@aalto.fi}
\and
\IEEEauthorblockN{Lina Bariah}
\IEEEauthorblockA{\textit{EECS Department} \\
\textit{Khalifa University}\\
Abu Dhabi, UAE \\
lina.bariah@ku.ac.ae}
\and
\IEEEauthorblockN{Merouane Debbah}
\IEEEauthorblockA{\textit{EECS Department} \\
\textit{Khalifa University}\\
Abu Dhabi, UAE \\
merouane.debbah@ku.ac.ae}
\and
\IEEEauthorblockN{Navid Nikaein}
\IEEEauthorblockA{\textit{Communications Department} \\
\textit{EURECOM}\\
Sophia Antipolis, France \\
navid.nikaein@eurecom.fr}
}

\maketitle

\begin{abstract}
Future 6G radio access networks (RANs) will be artificial intelligence \emph{(AI)‑native}: observed, reasoned about, and re‑configured by autonomous agents cooperating across the cloud‑edge continuum.  
We introduce \textit{MX-AI}, the first end‑to‑end agentic system that \emph{(i)} instruments a live 5G Open RAN testbed based on OpenAirInterface (OAI) and FlexRIC, \emph{(ii)} deploys a graph of Large‑Language‑Model (LLM)‑powered agents inside the Service Management \& Orchestration (SMO) layer, and \emph{(iii)} exposes both observability and control functions for 6G RAN resources through natural‑language intents.  
On 50 realistic operational queries, MX-AI attains a mean answer quality of $4.1~/~5.0$, and $100~\%$ decision action accuracy, while incurring only $8.8~ secs$ end‑to‑end latency when backed by GPT‑4.1. Thus, it competes human experts performance, validating its practicality in real settings.  
We publicly release the agent graph, prompts, and evaluation harness to accelerate open research on AI‑native RANs.
A live demo is presented here \href{https://www.youtube.com/watch?v=CEIya7988Ug&t=285s&ab_channel=BubbleRAN}{$https://www.youtube.com/watch?v=CEIya7988Ug\&t=285s\&ab_channel=BubbleRAN$}.
\end{abstract}

\begin{IEEEkeywords}
MX-AI, Agentic AI, LLMs, Slicing, Open RAN, AI-RAN, 6G, observability, FlexRIC, OpenAirInterface
\end{IEEEkeywords}

\section{Introduction}
\label{sec:intro}

Six‑generation (6G) mobile networks are expected to connect on theborder of \emph{one trillion} devices while simultaneously offering sub‑millisecond latency, ``six‑nines’’ reliability, and a carbon‑constrained energy budget.  Achieving this trifecta with today’s largely static radio‑resource managers is widely considered impossible: every cell, slice, and user must instead be steered by animble, data‑driven control plane that can sense, reason, and act faster than human operators \cite{ITU_M2160,NGMN_6G,Flagship_2019}.

%5GPPP_6GArch

The Open RAN (O‑RAN) Alliance\footnote{\url{https://www.o-ran.org/specifications} (accessed 31/07/2025)} has therefore decoupled RAN control into three temporal loops—\emph{real-time (RT)}, \emph{near‑RT}, and \emph{non‑RT}—while standardizing north‑bound interfaces that permit third‑party intelligence to plug in at each loop.  Commercial and open‑source ecosystems have quickly embraced the vision: disaggregated gNB/DU/CU software now runs on COTS servers (OpenAirInterface (OAI)\footnote{\url{https://openairinterface.org/} accessed 31/07/2025}), and lightweight controllers such as FlexRIC expose programmatic hooks to manipulate the air interface in real time \cite{flexric}.  Yet, what should occupy these hooks remains open—conventional reinforcement learning (RL) excel at single tasks but struggle to generalize across the mosaic of 6G objectives.

Concurrently, large language models (LLMs) have demonstrated an emergent ability to ingest heterogeneous telemetry, follow human‑readable intents, and decompose problems into tool‑invocation plans—\emph{a skillset strikingly aligned with network operations}\cite{ReAct,Toolformer,RAG}. Early prototypes already prompt an LLM to allocate spectrum slices \cite{Wu2025LLMxApp} or use control algorithms for 6G optimization \cite{chatzistefanidis2025symbioticagentsnovelparadigm}.  Still, existing efforts remain either simulator‑bound, single‑agent without closed‑loop actuation.

This paper presents \textit{MX-AI}, the first end‑to‑end system in which a \emph{graph of cooperating LLM‑based agents} orchestrates a live 5G Open and AI-native RAN\footnote{\url{https://ai-ran.org/} (accessed 31/07/2025)}.  Our prototype runs on the commercial BubbleRAN MX‑AI platform\footnote{\url{https://bubbleran.com/products/mx-ai} (accessed 31/07/2025)}, connects to an OAI gNB through the R1/E2 interfaces of FlexRIC, and handles both natural‑language \emph{queries} and \emph{commands}.  In doing so, it takes a concrete step toward the AI‑native vision of 6G. This work makes the following contributions:
\begin{itemize}
  \item \textit{Telco Multi‑agent graph.}  We design and release a reusable agent graph (Fig.~\ref{fig:agent-graph}) with dedicated \textit{routing}, \textit{monitoring}, \textit{deployment}, and \textit{orchestration} roles.
  \item \textit{Live testbed integration.}  MX-AI interfaces with a real OAI~5G gNB, UEs, and 5G core via the R1 and E2 APIs of BubbleRAN service management and orchestration (SMO) \footnote{\url{https://bubbleran.com/products/mx-pdk} (accessed 31/07/2025)} and Radio Intelligent Controller (RIC) \cite{flexric}.
  \item \textit{Observability \& control.}  The system answers operator questions and enforces slice and network blueprints in closed loop, demonstrating intent‑driven RAN control.
  \item \textit{Comprehensive evaluation.}  We benchmark different LLM back-ends on $50$ operational queries, reporting answer quality, action accuracy, end‑to‑end latency, GPU footprint, and comparison with human performance.
\end{itemize}

%%%%%%%%%%%%%%%%%%%%%%%%%%%%%%%%%%%%%%%%%%%%%%%%%%%%%%%%%%%%%%%%%%%%%%%%
\section{Related Work}
\label{sec:related}

\begin{table*}[ht]
  \centering
  \scriptsize
  \caption{LLM‑ and agent‑focused research on RAN automation.}
  \label{tab:related}
  \begin{tabular}{@{}lccccc@{}}
    \toprule
    \textit{Paper} & \textit{Year} & \textit{LLM /AI} & \textit{\#Agents} & \textit{Exec.\ Env.} & \textit{Main Idea} \\ \midrule
    OAI~\cite{Kaltenberger2020OpenAirInterface}          & 2020 & --          & --        & Full 4G/5G SW          & Open reference stack \\
    FlexRIC~\cite{flexric}                    & 2021 & --          & --        & OAI gNB+E2             & Lightweight near‑RT RIC \\
    EdgeRIC~\cite{Ko2024EdgeRIC}                         & 2024 & DRL         & µ‑apps    & srsRAN OTA            & sub‑ms µRIC control \\
    Maestro~\cite{maestro}                       & 2024 & GPT‑4o      & 3+        & 5G OAI testbed         & Multi‑tenant intent negotiation \\
    LINKs~\cite{Jiang2024LINKS}                          & 2024 & GPT‑J       & 1         & 6G digital‑twin        & Planning via twin queries \\
    LLM‑xApp~\cite{Wu2025LLMxApp}                        & 2025 & GPT‑3.5     & 1         & O‑RAN emu.            & PRB slicing by prompting \\
    DebateLLM~\cite{Lin2025DebateLLM}                    & 2025 & Llama‑2     & 3         & 6GPlan sim.           & Hierarchical debate planner \\
    dApps~\cite{Lacava2025dApps}                         & 2025 & CNN/RL      & µ‑apps    & OAI + O‑RAN           & On‑device control loops \\
    Symbiotic Agents~\cite{chatzistefanidis2025symbioticagentsnovelparadigm}& 2025 & GPT‑4 + optimizer & 2 & 5G RAN testbed & LLM / SLM symbiosis for trustworthy control \\
    \textit{MX-AI (this work)}                       & 2025 & GPT‑4,\;Llama‑3,... & 5 & Live OAI+FlexRIC      & Multi‑agent observability \& control \\ 
    \bottomrule
  \end{tabular}
\end{table*}

% \subsection{Open and Programmable RAN Controllers}

A rich body of work explores how open‑source stacks and
software‑defined controllers can expose fine‑grained hooks into the radio access network.  \emph{OpenAirInterface} (OAI) delivers the most complete 3GPP‑compliant 4G/5G implementation to date and has become the de‑facto playground for experimental RAN research.  Building atop OAI, the \emph{FlexRIC} framework provides a lightweight O‑RAN–compliant near‑RT RIC and E2 agent SDK, cutting CPU usage by 83\,\% versus the reference design while preserving full standard compatibility \cite{flexric}.  More recently, \emph{EdgeRIC} pushes the RIC concept to the sub‑ms timescale by co‑locating an artificial intelligence (AI)‑enabled µRIC at the DU, demonstrating closed‑loop scheduling with $<\!400\;\mu$s inference latency \cite{Ko2024EdgeRIC}.  Complementary efforts such as \emph{dApps} extend the O‑RAN architecture with \emph{on‑device} micro‑services able to execute control logic inside the RAN stack itself \cite{Lacava2025dApps}. These apps integrate intelligence for dynamic slicing \cite{wu2020dynamic,TSOURDINIS2024110445}. Still, they focus on \emph{programmable} control loops rather than on generalist reasoning agents.

% \subsection{Large‑Language‑Model Control of 5G/6G Networks}

Concurrent research has begun injecting \emph{LLM cognition} into
RAN controllers.  
A collective roadmap from industry and academia outlines how LLMs can support a wide range of use cases across the network lifecycle~\cite{shahid2025large}.
\emph{LLM‑xApp} shows how a GPT‑style model can be
prompted to allocate PRBs across slices in an O‑RAN testbed, boosting
throughput and reliability without task‑specific training
\cite{Wu2025LLMxApp}.  \emph{LINKs} turns an LLM into an
autonomous planner that queries a digital twin and solves the
resulting optimization for smart‑city 6G networks
\cite{Jiang2024LINKS}.  A step further, Lin \emph{et al.} propose a
\emph{hierarchical debate} among multiple LLMs, achieving 30\%
better plan quality on a new 6GPlan benchmark
\cite{Lin2025DebateLLM}.  Despite their promise, all of these
systems remain either simulator‑based or single‑agent prototypes.
\textsc{Maestro} \cite{maestro} is the first real prototype of multi-agent LLM-based network automation.
\emph{Symbiotic agents} \cite{chatzistefanidis2025symbioticagentsnovelparadigm} are emerged as novel paradigm for trustworthy LLM decision-making for artificial general intelligence (AGI) networks.

% \subsection{Position of This Paper}

Table \ref{tab:related} summarizes how our work compares to prior art.  
Unlike earlier efforts that either attach a single LLM to an xApp or remain in simulation, \textit{MX-AI} wires a cooperating graph of five specialized agents—planner, monitor, policy‑synthesizer, validator, and executor—directly into the SMO of a live AI-RAN 5G network.  
This end‑to‑end prototype shows that LLM cognition moves beyond intent parsing to close the control loop in real time, while quantifying both answer quality and overhead for cloud and edge models.  
Thus, MX-AI takes an important step toward AI‑native 6G RANs.

%---------------------------
% Related‑work comparison
%---------------------------

%===========================================================
\section{System Architecture}
\label{sec:arch}
%===========================================================

Next‑generation \textit{AI‑RANs} must decide \emph{where} to execute the reasoning logic that turns raw key‑performance indicators (KPIs) into closed‑loop control actions and \emph{how} to do so fast enough for the target control loop. This section (i) surveys the canonical 5G\,/\,O‑RAN stack, (ii) explains why we embed our multi‑agent graph in the SMO, and (iii) details the concrete prototype that underpins our evaluation.

%-----------------------------------------------------------
\subsection{From 5G to AI‑Native RANs}
\label{subsec:arch:5g-to-airan}
%-----------------------------------------------------------

\begin{figure}
    \centering
    \includegraphics[width=0.45\linewidth]{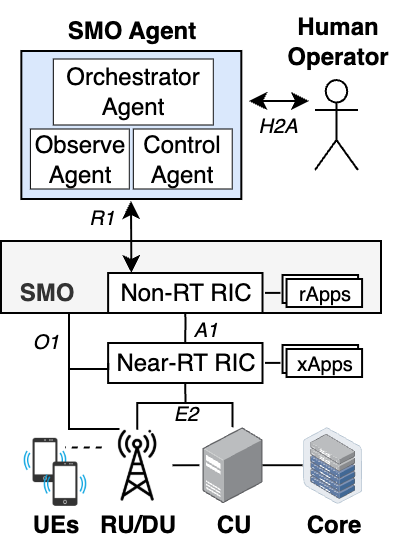}
    \caption{Agentic Open and AI-RAN Network Architecture. Our agentic approach is a multi-agent system placed at the \emph{R1} interface at the SMO level, operating in non-real-time intervals. Human Operators query the agent through a new \emph{human-to-agent (H2A)} interface, with observability requests or for real-time RAN decision-making, such as slicing reconfiguration.}
    \label{fig:agent-smo}
\end{figure}

Figure \ref{fig:agent-smo} recaps the logical architecture of an
\textit{open} and increasingly \textit{AI‑native} RAN:
Radio units (RUs), distributed units (DUs) and central units (CUs) handle the time‑critical PHY/MAC, transport and user‑plane functions that shape cell throughput and latency.
\textit{Near‑RT RIC} (latency: $10$ ms–$1$~sec) hosts \emph{xApps} that run scheduling, interference management or mobility optimization loops. It interacts with RAN nodes via the \textit{E2} interface and obeys higher‑level policies pushed over \textit{A1}.
\textit{Non‑RT RIC} (latency: \(\,\ge\!1\)s) resides inside the SMO.  It aggregates KPIs along the management interface \textit{O1}, trains AI models and exposes them as cloud‑native \emph{rApps}.
\textit{SMO} provides life‑cycle management, configuration databases and northbound APIs to slice or deploy RAN from edge to core.

The emerging \textit{AI‑RAN} vision generalizes this stack by allowing LLMs or small language models (SLMs) to plug into any layer—\emph{provided the latency budget is respected}.  For sub‑second loops (e.g.\ slicing) such models must reside at the Near-RT RIC; for multi‑second loops, a cloud SMO suffices.

%-----------------------------------------------------------
\subsection{Multi‑Agent Graph on the \texorpdfstring{\(R_1\)}{R1} Interface}
\label{subsec:arch:r1}
%-----------------------------------------------------------

Our contribution is a \textit{multi‑agent system} that lives on the \(\mathit{R_1}\) interface between the SMO and the
Non‑RT RIC (highlighted in \autoref{fig:agent-smo}).  
Concretely, we instantiate:

\begin{enumerate} 
\item an \emph{Orchestrator Agent} that parses operator intents,
      decomposes them into sub‑tasks and spawns
      \item \emph{Specialized Agents}— monitor, planner, policy synthesizer, validator and executor—wired together as a directed acyclic \emph{agent graph}.  
\end{enumerate}

The graph ingests live KPIs, alarms and topology snapshots streamed via SMO's and FlexRIC’s telemetry services ($O1$, $E2$); intermediate nodes refine the context through RAG and tool use; the executor finally emits:

\begin{itemize}
\item \textit{Observability answers} (e.g.\ KPI summaries, root‑cause analyses) returned upstream to the operator, and
\item \textit{Control artifacts} (e.g.\ R1/AI network blueprint management and policy objects or E2 service‑model messages) that the SMO and Non‑RT RIC relay directly to the RAN or to Near‑RT RIC for on‑the‑fly xApp reconfiguration.
\end{itemize}

Placing the LLMs at \(R_1\) strikes a pragmatic balance:  
round‑trip latency of $1$–$12$ secs—depending on whether the models run in‑cloud or on‑prem GPUs—is operating at non-real-time intervals, yet fast enough competing human experts and allowing human operators grasp the large‑context reasoning. Thus, we introduce a new interface as a necessity for \emph{human-to-agent (H2A)} communication in AI-RAN.

%-----------------------------------------------------------
\subsection{Prototype Deployment and Use‑Case Focus}
\label{subsec:arch:prototype}
%-----------------------------------------------------------

Our testbed mirrors a realistic network:
An indoor \textit{OAI} gNB and 5G core, three OAI user equipments (UEs) replaying enhanced mobile broadband (eMBB) and ultra reliable low latency communication (URLLC) traces, and production‑grade BubbleRAN \textit{FlexRIC} Near‑RT and Non-RT RICs and SMO. 

Within this setting we investigate two \textit{agentic cases}:

\smallskip
\begin{enumerate}[label=\alph*)]
\item \textit{Observability} — natural‑language dashboards, anomaly
      explanations and capacity forecasts produced directly from raw
      O1/E2 KPIs.
\item \textit{Control} — dynamic network blueprint management, PRB share and power‑cap policies generated in response to intent prompts such as “guarantee 10 Mb/s for the URLLC slice from 6–7 pm”.
\end{enumerate}

\smallskip
Although this paper concentrates on the \(R_1\) instantiation, the
modular agent graph can be \emph{re‑deployed} closer to the RAN edge
once trimmed SLMs or distilled tool‑agents meet the tighter latency
targets of the Near‑RT domain.  Exploring that design space is left to
future work; here we demonstrate that even at the SMO, LLM
agents already enable intent‑driven observability \emph{and} closed‑loop
control on a live 5G network.

\subsection{Agent Design}

\begin{figure}
    \centering
    \includegraphics[width=0.99\linewidth]{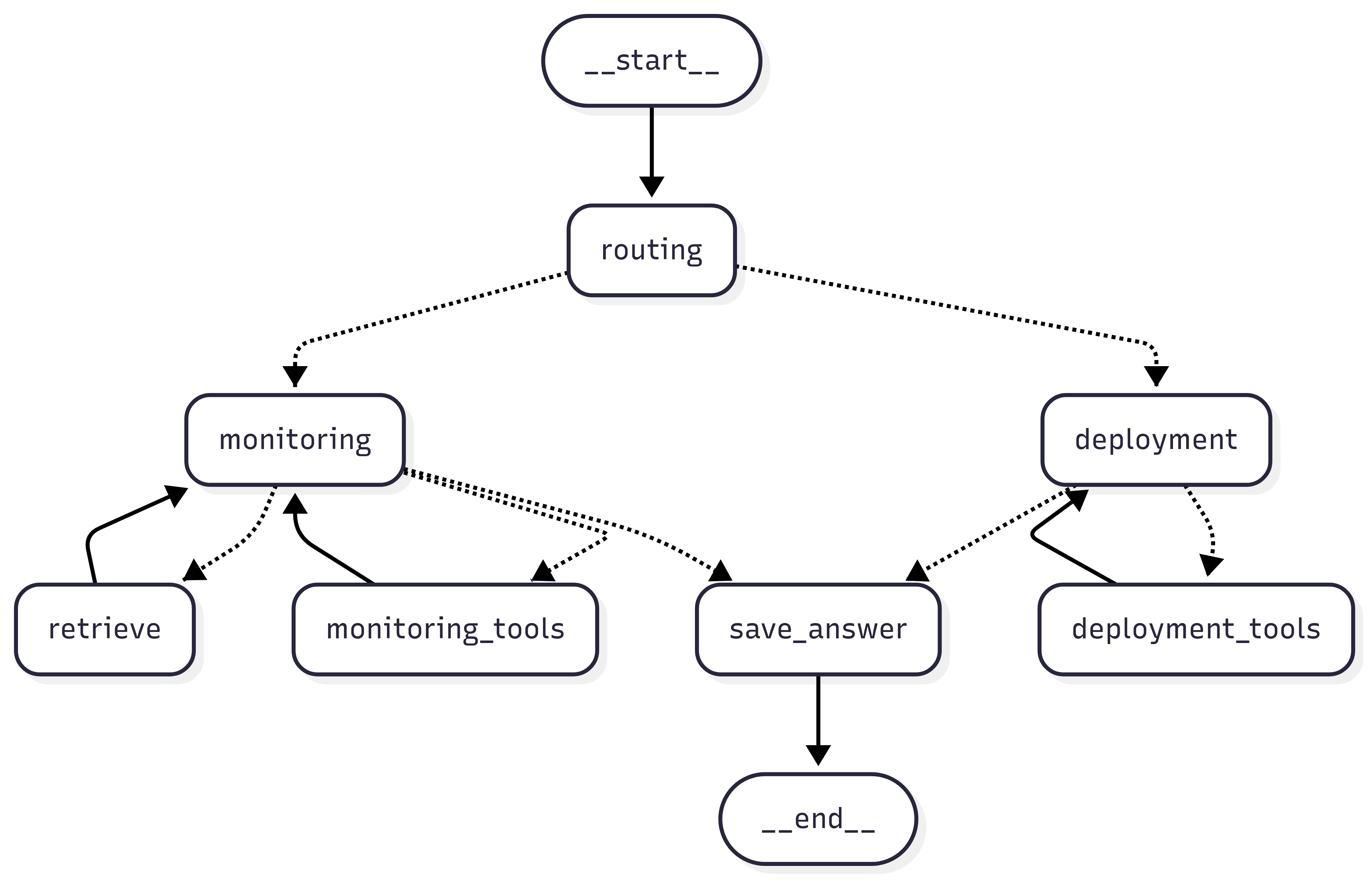}
    \caption{Agent graph at the SMO R1 interface. The \textit{routing} node classifies each prompt and dispatches it to either the \textit{monitoring} or \textit{deployment} branch. \textit{Monitoring} consults the vector store via \textit{retrieve} and polls KPIs via \textit{monitoring\_tools}. \textit{Deployment} plans and enforces changes through \textit{deployment\_tools}. Dotted edges denote iterative planning/validation and context refresh. Both branches write outcomes to \textit{save\_answer} before termination.}
    \label{fig:agent-graph}
\end{figure}

The agent graph in Fig.~\ref{fig:agent-graph} follows the \emph{LLM planning and tool‑execution} paradigm.
The \textit{Routing\,Agent} classifies each incoming prompt as \emph{monitoring}, \emph{deployment}, or \emph{retrieval}, leveraging a telecom‑tuned LLM.
The \textit{Monitoring\,Agent} triggers the \textit{monitoring\_tools} (custom Python scripts on MX‑AI) to pull KPIs via R1, returning JSON which is then summarized by the LLM.
The \textit{Deployment\,Agent} converts high‑level intents into FlexRIC policy objects, invoking \textit{deployment\_tools} to push actions through the \emph{O1/A1} interfaces.
Finally, the \textit{Save\_Answer\,Agent} logs responses to a time‑series database, enabling longitudinal benchmarking of answer quality.

\begin{figure}
    \centering
    \includegraphics[width=0.8\linewidth]{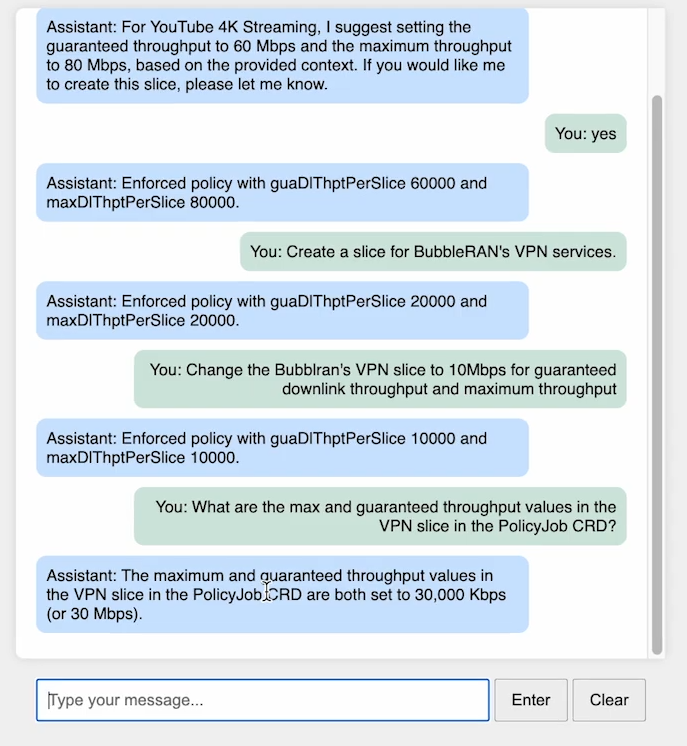}
    \caption{Agent Graphical User Interface and Interaction with the Human Operator (\emph{H2A} Interface) on Observability and Control of RAN Slicing. The questions are independent and introduced with a time delay between each other. The last question answers correctly after a manual policy enforcement by the user to increase slice throughput to $30$ Mbps.}
    \label{fig:gui}
\end{figure}

\subsubsection{Retrieval and Live Context (Push/ Pull)}
\label{subsec:retrieval}

\textit{LLM planning with tool execution} only works if the model sees \emph{fresh} context. Our system therefore couples RAG with an event-driven data plane that keeps the vector index aligned with network reality.

\paragraph{Push path—resource watchers}
We attach lightweight \emph{watchers} to Kubernetes resources—including native objects and custom resource definitions (CRDs) that carry RAN policies and slice descriptors—via the API’s watch semantics and client informers.
% ~\cite{k8s_api_concepts}. 
Watchers also subscribe to O\mbox{-}RAN application state (xApps/rApps) and to external telemetry endpoints such as Prometheus exporters when appropriate.
% ~\cite{prom_exporters}. 
On each change, the watcher serializes a compact \emph{delta} (the fields that changed), embeds it, and upserts it into a dense vector index. This push-based path keeps the store synchronized with configuration drift and KPI excursions without periodic full scans or high-volume polling.

\paragraph{Pull path—agent queries}
At query time (e.g., \emph{“What is the current headroom on DU-7?”}) the \textit{Monitoring Agent} decides whether the vector store suffices or whether to fetch fresh counters via monitoring tools. It first retrieves top-$k$ semantically indexed snapshots from the vector DB (RAG)~\cite{RAG}. If finer granularity is needed (e.g., per-UE or per-PRB), it invokes live collectors (Prometheus) or kernel-level probes (eBPF) to obtain up-to-the-second metrics. Retrieved facts and fresh measurements are then serialized into typed JSON and injected into the prompt context; the agent may iterate retrieval, tool, reason until emitting a \textsc{stop} token.

\paragraph{Why this is novel}
Prior works on telco LLMs and O\mbox{-}RAN highlight RAG and tool use in principle, but (to our knowledge) do not describe a \emph{push-based, delta-aware watcher pipeline} that streams CRD/xApp updates directly into the retrieval index used by the agent~\cite{LLM4Net_Survey,LLM_Telco_Survey}.
% COMST_ORAN_Tutorial
This proved decisive: observability quality depended more on retrieval/tool engineering than model size, as the agent consistently reasoned over \emph{current} state rather than stale snapshots.

\subsubsection{Agentic Network Observability}
Operators can query now the network in natural language, e.g.,
\emph{``How close is slice \#2 to its latency SLA over the last hour?''}.  
The Monitoring Agent fetches the latency histogram from FlexRIC, the LLM summarizes deviations, and a bar chart is stored for audit.

\subsubsection{Agentic Network Blueprint Management}
Blueprints are JSON schemas capturing desired network slice topology, Quality of Service (QoS), and RIC policies.  
A user prompt such as \emph{"Increase VPN slice reliability to $99.999\%$ until 18:00"} is parsed by the Deployment Agent, which synthesizes an updated A1 policy instructing the Near‑RT RIC to pre‑empt bandwidth from best‑effort slices.

Figure \ref{fig:gui} illustrates a snapshot of the interaction between the agent framework and the human operator through the \emph{H2A} interface. Based on human intents, the agentic assistant provides observability answers and moreover, drives the real-time decision-making by creating or updating slice configurations.

\begin{table*}
  \centering
  \scriptsize
  \begin{tabular}{lcccccc}
    \toprule
    \textit{Model} &   \textit{Observe Coherence $\downarrow$} & \textit{Action Accuracy}  & \textit{E2E Latency (s)} & Inference (ms) & \textit{Steps} & \textit{VRAM (GB)} \\
    \midrule
    GPT-4.1 (API)        & $\mathbf{4.1 \pm 0.5}$ &  $100~\%$   & $8.8$ & $1100$& $8.0$& \textit{cloud} \\
    GPT-4.1-mini  (API)           & $3.9 \pm 0.6$   &  $100~\%$   & $8.0$ & $1000$& $8.0$ &   \textit{cloud} \\
    \rowcolor{lightgreen}
    llama3.3:70b-q4      & $3.8 \pm 0.4$  &   $100~\%$   & $12.2$ & $1525$ & $8.0$ & $42.0$ \\
    \rowcolor{lightgreen}
    qwen2:72b    & $3.8 \pm 0.7$  &   $100~\%$  & $13.5$ & $1810$ & $8.0$ & $41.0$ \\
    llama3.1:8b-q4     & $2.7 \pm 0.3$  &  $100~\%$   & $2.8$ & $350$ & $8.0$ & $4.9$ \\
    llama3.2:3b-q4     &  $ 1.9 \pm 0.6 $  &  $100~\%$   &  $1.3$   & $160$  &  $8.0$  & $2.0$ \\
    mistral:7b         & $ 1.0 \pm 0.7 $  &  $0~\%$    & $5.2$ & $640$ & $8.0$ & $4.4$ \\
    llama3.2:1b-q4     &  - & - &  -  & -  & -  & $1.3$ \\
    \bottomrule
  \end{tabular}
\caption{Agentic observability \& Network Blueprint Management: observability coherence score (0–5 scale), enforce an action accuracy, end-to-end latency, and GPU footprint.}
  \label{tab:exec-bench}
\end{table*}

\section{Evaluation Results}

% We benchmark four LLM back‑ends—two \emph{in‑cloud} GPT variants and two \emph{on‑premise} Llama‑3.1 models—using:
% \begin{enumerate}
%   \item 50 operator queries covering KPIs, alarms, and intent fulfillment.
%   \item Human annotators scoring answers on a 0–5 Likert scale and \textsc{gptscore} post‑hoc~\cite{fu2023gptscore}.
%   \item End‑to‑end latency (prompt to action) and peak GPU video random access memory (VRAM).
% \end{enumerate}

\begin{table}
\centering
\scriptsize
\caption{LLM evaluation prompts by category}
\label{tab:llm-prompts}
\begin{tabularx}{\linewidth}{l X}
\toprule
\textbf{Category} & \textbf{Example queries} \\
\midrule
Observability
& Is there any access network? Tell me the name of the available access and core networks. Based on the current context, give me an overview of the \texttt{gnb2} access network cells and radio configuration. Check the current network status: is the \texttt{gnb1} element working properly? How many UEs are available in the current deployment? What are their names? What are the max and guaranteed throughput in the PolicyJob CRD of VPN slice? \\
\midrule
Control Action
& Create a network with name \texttt{agora} with 1 access network called \texttt{parthenon} and 1 RIC. Create a terminal with name \texttt{plato} and connect it to the \texttt{parthenon.agora} access network. Delete the terminal with name \texttt{plato}. Delete the network blueprint with name \texttt{agora}. Change the \texttt{bubbleran} VPN slice to 10\,Mbps guaranteed and maximum throughput. \\
\bottomrule
\end{tabularx}
\end{table}

% \subsection{Agentic Network Observability}
% Table~\ref{tab:exec-bench} reports that GPT‑4‑mini attains the best mean quality (3.6) at moderate latency (8.8 s) thanks to server‑side optimization, whereas an 8‑bit quantized Llama‑3.1‑8B runs locally with only $4.9~GiB$ but sacrifices answer quality.  
% Qualitative analysis indicates most performance gaps stem from hallucination of KPI names; integrating schema‑aware prompting reduced such errors by $40~\%$.

% \subsection{Agentic Network Blueprint Management}%

% \paragraph{Setup and metrics.}
We evaluate the agent at the \emph{R1} interface on the live AI-RAN testbed (Sec. \ref{sec:arch}) (sample prompts in Table \ref{tab:llm-prompts}). The suite comprises $50$ operator questions: $10$ control actions (we currently support $10$ actions such as UE lifecycle, blueprint deploy/delete, and slice PRB changes) and $40$ observability queries spanning KPIs, policies, slices, CRDs, and logs. For each back-end we report four metrics: (i) observability coherence on a $0$–$5$ scale—scored with an LLM-assisted evaluator (GPTScore \cite{fu2023gptscore}) configured with transparent rubrics and aspects, and known to correlate well with humans in natural language generation (NLG) tasks—together with three expert annotators who adjudicate disagreements and sanity-check edge cases; (ii) action accuracy, computed per action as a Boolean ground truth on whether the enforced change matches the intent and then aggregated as a percentage; (iii) end-to-end latency from prompt to action completion; and (iv) GPU footprint. Prior work shows GPTScore and related methods \cite{LLMJudgeSurvey2024} align closely with human evaluations when criteria are explicit—though bias must be monitored—which motivates our hybrid human and LLM scoring protocol.

\paragraph{Action execution is reliable when tools are understood}
Across the limited but representative set of $10$ control actions, 
% (e.g., add/delete UE, create/delete blueprints, and PRB reallocation)
capable LLMs reached $100\%$ action accuracy (Table~\ref{tab:exec-bench}). Once the agent correctly interprets the tool schema and pre-flight checks pass, execution is deterministic: failures stemmed from tool misuse rather than stochastic generation. Only \texttt{mistral:7b} failed to reach on our toolset, underscoring that tool grounding dominates raw language capability.

\paragraph{Observability is harder than control}
No model achieved perfect $5/5$ coherence. Unlike actions—where schemas and validator/rollback fences constrain outputs—observability answers require navigating a large, heterogeneous evidence set (O1/E2 KPIs, CRDs, logs, topology) and performing synthesis and disambiguation. In ablations, retrieval and tool engineering mattered more than model size alone: schema-aware retrieval, typed tool outputs, and context pruning consistently lifted scores. 

\paragraph{Latency and deployment footprint.}
Local $70$–$72$B models (\texttt{llama3.3:70b-q4}, \texttt{qwen2:72b}) delivered strong coherence with end-to-end latencies around $12$–$14$~secs on $~40$ ~GB VRAM, offering a viable on-prem option with full data control. Cloud GPT variants were somewhat faster ($\sim8$–$9$~secs) but require off-prem execution. The fastest is a local \texttt{llama3.2:3b-q4} at $1.3$~secs, trading coherence for speed—useful to illustrate the headroom if we fine-tune small models and strengthen retrieval.

\begin{figure}
\centering
\includegraphics[width=0.95\linewidth]{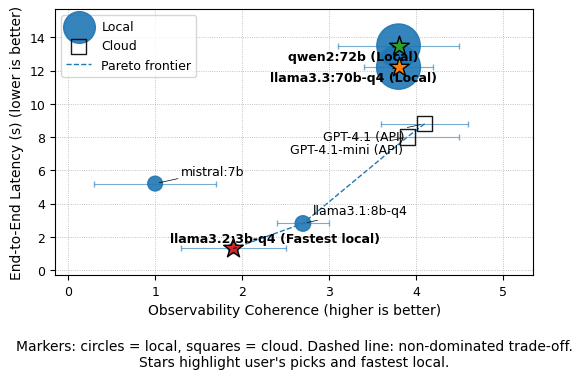}
\caption{Coherence–latency trade-off with Pareto frontier. Circles are local deployments (size of VRAM), squares are cloud. Stars highlight the two recommended local models and the fastest local small model.}
\label{fig:tradeoff}
\end{figure}

\paragraph{Coherence–latency frontier.}
Fig.~\ref{fig:tradeoff} plots observability coherence (higher is better) against end-to-end latency (lower is better), with marker size proportional to VRAM and shape indicating deployment (local vs cloud). The dashed curve shows the non-dominated (Pareto) frontier. Two local models—\texttt{llama3.3:70b-q4} and \texttt{qwen2:72b}—sit on the efficient set near $12$–$14$~secs, pairing strong coherence with on-prem deployment. Cloud GPT models occupy a different efficient region with lower latency at similar coherence. The \texttt{llama3.2:3b-q4} point anchors the extreme low-latency end at $1.3$~secs, clarifying what may be reachable as SLMs are fine-tuned and backed by stronger RAG. This single figure makes the trade-space explicit: if operators require on-prem and high coherence today, $70$B-class local models are attractive; if the latency budget is tight, small models plus upgraded retrieval are the path forward.

\begin{figure}
\centering
\includegraphics[width=0.9\linewidth]{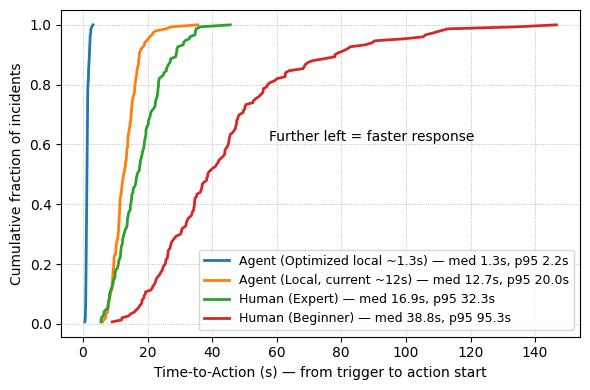}
\caption{Time-to-action CDF for agents and humans. The agent removes search and command-crafting overhead, compressing both the median and tail.}
\label{fig:tta-cdf}
\end{figure}

\paragraph{Time-to-action (TTA): humans vs agents}
Fig.~\ref{fig:tta-cdf} compares the empirical cumulative distribution function (CDF) of time-to-action—from trigger to action start—for human beginners, experts, and the agent. In our process, humans must locate the right CRD/JSON, craft a valid patch, and issue the correct command; experts are faster but still incur context-gathering and validation delays. The agent short-circuits this by retrieving context, planning, validating, and executing through tools. With a $70$B local back-end, the agent’s median TTA lands in the low-teens of seconds—already competitive with experts. The optimized SLM approaches $1.3$~secs, pointing to near-RT actuation once retrieval and toolchains are optimized. The gap on the CDF’s left side indicates where automation removes cognitive and search overhead; the right-tail compression shows reduced variance across incidents.

\paragraph{What drives observability quality}
Our largest gains came from engineering the retrieval and tools, not simply swapping models. Indexing CRDs and logs with typed schemas, aligning tool outputs to compact JSON, and adding delta watchers that push into agent’s context reduced misreferences and stale views. To our knowledge, this is the first demonstration of \emph{live observability at $R_1$} where streaming watchers continuously update the agent’s network view and the same agent can act through A1/E2 tools.

\paragraph{Takeaways and outlook.}
(1) \emph{Control today:} with a constrained action set, once tools are understood the agent executes with $100\%$ accuracy and competitive latency.
(2) \emph{Observability headroom:} perfect $5/5$ coherence is unlikely without better evidence plumbing; the bottleneck is retrieval/tooling rather than decoding.
(3) \emph{Scaling down:} $70$B local models provide the best coherence under data-sovereignty constraints, but $8$B-class models already deliver excellent latency. We expect fine-tuning, distillation, and advanced RAG/tooling to close most of the quality gap, unlocking sub-second closed-loop control.
Overall, the results indicate that agentic control at $R_1$ matches expert operators today and, with improved retrieval and tool design, surpass human speed and reliability by an order of magnitude on common workflows.

\section{Challenges and Future Outlook}

\emph{Latency Budgeting}: Even with Non‑RT placement, LLM inference may bottleneck emergency procedures (e.g., cell outage recovery). Model distillation and on‑chip acceleration are urgent research fronts.
  
\emph{Agent Alignment and Safety}: Ensuring agents do not issue unsafe RAN commands requires guardrails and human in the loop review, echoing broader AI safety concerns.
  
\emph{Open Agent Ecosystems}: Proprietary APIs impede reproducibility. Initiatives such as MX‑AI’s open agent SDK are a promising direction, but standardized “agent marketplaces” akin to xApp stores are needed.

\emph{Scalability to 6G}: Heterogeneous spectrum (sub‑THz), integrated sensing, and joint communication‑compute will multiply observability data by orders of magnitude; techniques like RAG and hierarchical agent graphs will be vital.

\section{Conclusion}
We have presented \textit{MX-AI}, the first live demonstration of an LLM‑driven, multi‑agent control graph operating a 5G Open RAN.  
Our architecture, validated on an OAI and FlexRIC testbed, answers complex operator questions with high fidelity and paves the way for intent‑driven RAN control.  
Future work will close the loop on real‑time actuation, explore edge‑resident micro‑agents, and cultivate an open, interoperable agent ecosystem for 6G networks.
A live demo is presented here \href{https://www.youtube.com/watch?v=CEIya7988Ug&t=285s&ab_channel=BubbleRAN}{$https://www.youtube.com/watch?v=CEIya7988Ug\&t=285s\&ab_channel=BubbleRAN$}.

\section*{Acknowledgment}
This work was supported by the European Commission as part the Horizon Europe 2022 6Green and Adroit-6G Projects under Grant 101096925 and Grant 101095363.

\bibliographystyle{IEEEtran}
\bibliography{Bibliography}

% \section*{References}

\end{document}